\documentclass[10pt,aps,prl,twocolumn,twoside,superscriptaddress,nofootinbib,preprintnumbers,floatfix,longbibliography]{revtex4-2}
\usepackage{hyperref,graphicx,amsmath,xspace,booktabs,xcolor,todonotes}

\usepackage{mathtools}
\newcommand{\done}{{\rm d}}

\newcommand{\qT}{\ensuremath{q_\mathrm{T}}}
\newcommand{\qTvec}{\ensuremath{\vec{q}_\mathrm{T}}}

\newcommand{\bT}{\ensuremath{b_\mathrm{T}}}
\newcommand{\bTvec}{\ensuremath{\vec{b}_\mathrm{T}}}

\newcommand{\MCFM}{M\protect\scalebox{0.8}{CFM}\xspace}
\newcommand{\OpenLoops}{O\protect\scalebox{0.8}{PEN}L\protect\scalebox{0.8}{OOPS}\xspace}

\newcommand{\GeV}{\:\mathrm{GeV}}
\newcommand{\TeV}{\:\mathrm{TeV}}

\newcommand{\rT}{r_T}
\newcommand{\df}{\mathrm{d}}

\newcommand{\mll}{m_{\ell\ell}}
\newcommand{\dphi}{\Delta\phi_{\ell\ell}}
\newcommand{\acop}{\pi-\dphi}
\newcommand{\aco}{\phi_{\rm acop}}
\newcommand{\phistar}{\phi^*_\eta}
\newcommand{\Neff}{N_\mathrm{eff}}

\newcommand{\NkLL}{\mathrm{N}^4\mathrm{LL}^\prime}
\newcommand{\PWGPRIORDEV}{$17\%$}

\newcommand{\PWGNEFF}{$99\%$}
\newcommand{\NkLO}{\mathrm{N}^3\mathrm{LO}}

\newcommand{\eq}[1]{Eq.~\eqref{eq:#1}}
\newcommand{\eqs}[2]{Eqs.~\eqref{eq:#1} and~\eqref{eq:#2}}
\newcommand{\fig}[1]{Fig.~\ref{fig:#1}}

\newcommand{\prlsec}[1]{\par\medskip\noindent\textit{#1.}---}

\hypersetup{hidelinks,
  pdfauthor={Benoit Assi,Enrico Bothmann,John Campbell,Christian Gutschow,Stefan Hoeche,Wan-Li Ju,Max Knobbe,Marek Schoenherr,Jesse Thaler,Michael Wagman},
  pdftitle={Drell--Yan lepton pair production from low to high transverse momentum},
  pdfkeywords={QCD, Resummation, Lattice QCD, Monte Carlo, Drell-Yan}
}

\begin{document}
\preprint{FERMILAB-PUB-26-0706-T, IPPP/26/68, MIT-CTP 6120, MCNET-26-28}
\title{Drell--Yan lepton pair production from low to high transverse momentum}

\author{Beno\^{i}t Assi}
\email{bassi@perimeterinstitute.ca}
\affiliation{Perimeter Institute for Theoretical Physics, Waterloo, N2L 2Y5, Canada}
\affiliation{Department of Physics, University of Toronto and Vector Institute, Toronto, ON, M5S 1A7, Canada}
\author{Enrico Bothmann}
\affiliation{IT Department, CERN, CH-1211 Geneva, Switzerland}
\author{John Campbell}
\affiliation{Theory Division, Fermilab, Batavia, IL 60510, USA}
\author{Christian G{\"u}tschow}
\affiliation{Centre for Advanced Research Computing, University College London, Gower Street, London, WC1E 6BT, UK}
\author{Stefan H{\"o}che}
\email{shoeche@fnal.gov}
\affiliation{Theory Division, Fermilab, Batavia, IL 60510, USA}
\author{Wan-Li Ju}
\affiliation{Department of Physics, University of Alberta, Edmonton AB T6G 2J1, Canada}
\author{Max Knobbe}
\affiliation{Theory Division, Fermilab, Batavia, IL 60510, USA}
\author{Marek Sch{\"o}nherr}
\affiliation{Institute for Particle Physics Phenomenology, Durham University, Durham DH1 3LE, UK}
\author{Jesse Thaler}
\affiliation{Center for Theoretical Physics -- a Leinweber Institute, Massachusetts Institute of Technology, Cambridge, MA 02139, USA}
\affiliation{The NSF Institute for Artificial Intelligence and Fundamental Interactions}
\author{Michael L.~Wagman}
\affiliation{Theory Division, Fermilab, Batavia, IL 60510, USA}

\begin{abstract}
We present a technique for computing Drell--Yan lepton pair production
at particle-level that achieves N$^3$LO and approximate N$^4$LL$'$ accuracy at small to
moderate transverse momentum, tree-level precision for $Z+4$, $5$, and $6$~jets, and approximate
NLO EW precision at high transverse momentum. A crucial non-perturbative input,
the Collins--Soper kernel, is obtained from continuum-limit lattice QCD results.
High-precision theory results are recast into strictly positive event weights
using an information-theoretic approach based on the maximum entropy principle.
\end{abstract}
 \maketitle

The production of Drell--Yan lepton pairs in hadronic collisions has played
an important role from the early days of particle physics~\cite{Drell:1970wh,
  UA1:1983mne,UA2:1983mlz}. Properties of the final state, such as the lepton pair's transverse
momentum ($q_T$) and its angular distributions, have become precision
observables used both in the determination of Standard Model parameters and in searches
for new physics. Large corrections to the Drell--Yan production rate in perturbative
QCD~\cite{Altarelli:1978id,Altarelli:1979ub} have inspired higher-precision
calculations~\cite{Anastasiou:2003yy,Anastasiou:2003ds,Melnikov:2006kv,
  Li:2012wna,Campbell:2019dru,Duhr:2020sdp,Duhr:2020seh,Bonciani:2021zzf,
  Chen:2021vtu,Chen:2022cgv,Neumann:2022lft} as well as the development of resummation
techniques~\cite{Collins:1982wa,Collins:1983ju,Bodwin:1984hc,Collins:1985ue,
  Sterman:1986aj,Collins:1988ig} that have proved remarkably
successful~\cite{Bizon:2019zgf,AH:2020cok,Becher:2020ugp,Ebert:2020dfc,
  Camarda:2021ict,Re:2021con,Ju:2021lah,Billis:2024dqq}. The most precise
predictions today can be made up to approximate N$^4$LL$\times$N$^3$LO accuracy~\cite{
  Re:2021con,Chen:2022cgv,Neumann:2022lft,Camarda:2023dqn,Billis:2024dqq}.
They have improved the description of the intermediate and large transverse
momentum region, but the low transverse momentum domain has largely remained
subject to fitting. Recent developments in lattice QCD have changed this
picture by making the first non-perturbative component of the resummation,
the Collins--Soper kernel~\cite{Collins:1981uk,Collins:1981va,Collins:1984kg},
amenable to ab initio computation~\cite{Shanahan:2020zxr,LatticeParton:2020uhz,Li:2021wvl,Schlemmer:2021aij,Shanahan:2021tst,LatticePartonLPC:2022eev,Shu:2023cot,LatticePartonLPC:2023pdv,Avkhadiev:2023poz,Avkhadiev:2024mgd,Tan:2025ofx,Avkhadiev:2025wps}.

Experimental measurements of the Drell--Yan process rely on Monte-Carlo (MC) simulations,
which allow for the computation of observables under arbitrary fiducial cuts~\cite{
  Buckley:2011ms,Campbell:2022qmc}. This flexibility typically comes at the cost
of reduced precision compared to dedicated theoretical calculations.
NNLO-accurate generators~\cite{Karlberg:2014qua,Hoche:2014uhw,Alioli:2015toa,Monni:2019whf}
narrow this gap for Drell--Yan itself, but the
simulations employed by the experiments at the Large Hadron Collider (LHC) still
lag behind the precision of existing theory results.
In this Letter, we remove this limitation using a technique based on the
maximum entropy principle~\cite{Assi:2025ibi}.
We extend a state-of-the-art precision calculation with non-perturbative information
and transfer its results to simulations through a small number of moments of the
scaled transverse momentum $\rT=\qT/\mll$ and the lepton acoplanarity $\aco=\acop$.
Our theory prediction combines N$^3$LO~\cite{Boughezal:2015ded,Neumann:2022lft} with N$^4$LL$'$ resummation building on Ref.~\cite{Ju:2021lah}
and a Collins--Soper kernel from lattice QCD~\cite{Avkhadiev:2023poz,Avkhadiev:2024mgd}. The MC simulation uses
NLO QCD matching~\cite{Frixione:2002ik,Hoeche:2011fd} and multi-jet
merging~\cite{Hoeche:2009rj,Hoeche:2012yf} with approximate electroweak
corrections~\cite{Kallweit:2015dum,Lindert:2017olm}. The program that
derives event weights for any existing MC sample is publicly available~\cite{code}.

\prlsec{Lattice-improved $\qT$ resummation}
The Drell--Yan cross section, fully differential in the lepton
kinematics, can be written in the following form:
\begin{align}
\label{eq:qTres:mat}
\frac{\done\sigma_{\mathrm{DY}}}{\done\Phi_{\ell\bar\ell}}
=&
\frac{\done\sigma_{\mathrm{f.o.}}}{\done\Phi_{\ell\bar\ell}}
+f_{\mathrm{tran}}(\rT)
\left[
\frac{\done\sigma_{\mathrm{res}}}{\done\Phi_{\ell\bar\ell}}
-\frac{\done\sigma_{s}}{\done\Phi_{\ell\bar\ell}}
\right],
\end{align}
where \(\done\Phi_{\ell\bar\ell}=\done^2\qTvec\,\done Y_{\ell\bar\ell}\,\done m_{\ell\bar\ell}^2\,\done\Omega_{\ell\bar\ell}\)
is the di-lepton phase-space element,
\(\done\sigma_{\mathrm{f.o.}}\) is the fixed-order QCD expression, while
\(\done\sigma_{\mathrm{res}}\) is the resummed result and \(\done\sigma_s\) is its fixed-order expansion. 
The transition function \(f_{\mathrm{tran}}\) ensures the resummation is active at small \(\rT\) and
matches it to the fixed-order prediction smoothly as \(\rT\) increases (see End Matter).

Using soft-collinear effective theory (SCET) \cite{Bauer:2000yr,Bauer:2001yt,Beneke:2002ph},
the leading singular contribution in the low $\qT$ domain 
factorizes in impact parameter ($\vec{b}_T$) space as
 \cite{Becher:2010tm,Chiu:2011qc,Chiu:2012ir,Li:2016axz,Li:2016ctv}
\begin{align}
\label{eq:qTres:fac}
\frac{\done\sigma_{\mathrm{res}}}{\done\Phi_{\ell\bar\ell}}
=&
\int\frac{\done^2\bTvec}{(2\pi)^2}\,
e^{i\bTvec\cdot\qTvec}\,
\mathcal{U}^{\mathrm{s}}_{V}(\mu_b,\mu_s,\nu_b)\,
\mathcal{U}^{\mathrm{h}}_{V}(\mu_h,\mu_b,m_{\ell\bar\ell})
\nonumber\\
&\times
\left(\frac{\nu_b^2}{\nu_s^2}\right)^{
g_{\mathrm{pert}}(\mu_s,\mu_*)+g_{\mathrm{np}}}\,
\mathcal{H}(m_{\ell\bar\ell},\Omega_{\ell\bar\ell},
            \mu_t,\mu_h)\,
\nonumber\\
&\times
\mathcal{B}_{n}(\eta_n,\bTvec,\mu_b,\mu_f,\nu_b)\,
\mathcal{B}_{\bar n}(\eta_{\bar n},\bTvec,\mu_b,\mu_f,\nu_b)
\nonumber\\
&\times
\mathcal{S}_{\perp}(\bTvec,\mu_s,\nu_s)\,
\mathcal{S}_{\mathrm{mod}}(\bTvec)\,,
\end{align}
with
\(\eta_n=m_{\ell\bar\ell}e^{Y_{\ell\bar\ell}}/\sqrt{s}\) and
\(\eta_{\bar n}=m_{\ell\bar\ell}e^{-Y_{\ell\bar\ell}}/\sqrt{s}\).
The hard function $\mathcal H$ contains the short-distance virtual corrections extracted from the quark form factor in the heavy-top limit. The flavor-nonsinglet component is known through four loops~\cite{Moch:2005id,Moch:2005tm,Baikov:2009bg,Gehrmann:2010tu,Gehrmann:2010ue,Lee:2022nhh}, while the singlet component is available through three loops~\cite{Chen:2021rft,Gehrmann:2021ahy,Billis:2024dqq}. Top-induced axial-singlet effects are incorporated by the Wilson coefficient up to the fourth order~\cite{Chetyrkin:1993ug,Chetyrkin:1993hk,Chetyrkin:1993jm,Ju:2021lah,Chen:2021rft,Ahmed:2021spj,Chen:2021gxv,Chen:2022lun}.   
The soft function  $\mathcal{S}_{\perp}$  and the beam functions $\mathcal{B}_{n(\bar{n})}$ describe wide-angle soft and energetic collinear radiation, respectively. In the exponential rapidity-regulator scheme, both are known through N$^3$LO~\cite{Li:2016axz,Li:2016ctv,Luo:2019hmp,Luo:2019bmw,Luo:2019szz}.
The $\mu_f$ dependence in  $\mathcal{B}_{n(\bar{n})}$ is derived  up to the four-loop level by solving DGLAP equations~\cite{Moch:2004pa,Vogt:2004mw,Moch:2017uml,Davies:2016jie,Falcioni:2023luc,Falcioni:2023vqq,Gehrmann:2026qbl,Gehrmann:2023cqm,Moch:2023tdj}. 
Together with the requisite anomalous dimensions, these exact three-loop fixed-order inputs yield N$^3$LL$'$ resummation. For the approximate N$^4$LL$'$ extension, we include the four-loop logarithms fixed by the respective evolution equations and estimate the unknown soft and beam boundary terms in the large-$\beta_0$ approximation~\cite{Beneke:1994qe,Scimemi:2016ffw}.

Large virtuality logarithms are resummed by the hard and soft evolution
kernels, 
\begin{align}
\ln\mathcal{U}^{\mathrm{h}}_{V}
=&
\int_{\mu_h^2}^{\mu_b^2}\frac{\done\bar\mu^2}{\bar\mu^2}
\Bigg[
-\Gamma_{\mathrm{cusp}}
 \ln\!\bigg(\frac{\bar\mu^2}{m_{\ell\bar\ell}^2}\bigg)
+2\gamma_h
\Bigg] ,
\nonumber\\
\ln\mathcal{U}^{\mathrm{s}}_{V}
=&
\int_{\mu_s^2}^{\mu_b^2}\frac{\done\bar\mu^2}{\bar\mu^2}
\Bigg[\;
\Gamma_{\mathrm{cusp}}
 \ln\!\bigg(\frac{\bar\mu^2}{\nu_b^2}\bigg)
-\gamma_s
\Bigg] .
\label{eq:qTres:evolution}
\end{align}
The lightlike cusp anomalous dimension $\Gamma_{\mathrm{cusp}}$ is known exactly through four loops
\cite{Moch:2004pa,Moch:2018wjh,vonManteuffel:2020vjv}; at five loops we use
the estimate of Ref.~\cite{Herzog:2018kwj}.  The noncusp hard and soft
anomalous dimensions \(\gamma_h\) and \(\gamma_s\) are known through four loops
\cite{Lee:2022nhh,Duhr:2022cob,Gehrmann:2026qbl}.
The resummation of the rapidity logarithms is governed by the exponents
\(g_{\mathrm{pert}}\) and \(g_{\mathrm{np}}\):
  \begin{align}
  g_{\mathrm{pert}}
    =& \; \,       \gamma_r\left( \mu_*\right)+
          \int^{\mu_*^2}_{\mu^2_s}\frac{\done\bar{\mu}^2}{\bar{\mu}^2}\;
           \Gamma_{\mathrm{cusp}}\left[\alpha_s(\bar{\mu})
          \right] \,,\\
    g_{\mathrm{np}}=&  \;    -2\,  c_0\, \bT\,b_*\,,
    \quad b_*=\frac{\bT}{\sqrt{1+\bT^2/{B}_{\mathrm{NP}}^2}}\,. \label{eq:NPdef}
\end{align}
$g_{\mathrm{pert}}$  characterizes the perturbative contribution from
the small $\bT$ regime during the rapidity evolution, with the rapidity
anomalous dimension  $\gamma_r$ evaluated at  $\mu_*={b_0}/{b_*}$,
where $b_0=2e^{-\gamma_E}$.
The perturbative expression of  $\gamma_r$ up to four loops can be found
in Refs.~\cite{Duhr:2022cob,Duhr:2022yyp,Moult:2022xzt,Gehrmann:2026qbl}, while
$g_{\mathrm{np}}$  supplies the non-perturbative completion of the
Collins--Soper kernel, $g_{\mathrm{pert}}+g_{\mathrm{np}}$, at large \(\bT\).

Conventionally, the non-perturbative component is fitted to Drell--Yan data
and therefore not predictive. A key novelty of our analysis is that we extract $g_{\mathrm{np}}$ from continuum-extrapolated lattice
QCD~\cite{Avkhadiev:2023poz,Avkhadiev:2024mgd} with approximately physical quark masses and constraining the kernel over impact parameters
$0.15\,\mathrm{fm}\lesssim\bT\lesssim0.8\,\mathrm{fm}$ (the resummation-dominated
window $240\,\mathrm{MeV}\lesssim\qT\lesssim1.6\GeV$). For the
parametrization in Eq.~\eqref{eq:NPdef} with $B_{\mathrm{NP}}=2\GeV^{-1}$, lattice fits give
$c_0=0.032\pm0.012\GeV^2$~\cite{Avkhadiev:2024mgd}.

Nonperturbative corrections associated with the small-\(b_T\)
operator-product expansion of the beam functions are parametrized separately
as $ \mathcal S_{\mathrm{mod}}(\bTvec)=\exp\!\left(-{\bT^2}/{\kappa_{\mathrm{NP}}}\right)$. 
We employ the Gaussian form of Ref.~\cite{Billis:2024dqq} and calibrate its
width 
$\kappa_{\mathrm{NP}}=(3.0\pm0.5)~\GeV^{-2}$ from the \(8\) TeV~\cite{ATLAS:2015iiu} and \(13\) TeV ATLAS data
\cite{ATLAS:2019zci} in the low \qT\ region.
This width is the only non-perturbative parameter fitted to Drell--Yan data.
In the future, precise results will be available from lattice QCD for these nonperturbative effects~\cite{LatticePartonCollaborationLPC:2022myp}.

  The theoretical uncertainty consists of three components. First, resummation uncertainties are estimated by varying the auxiliary scales about the central choices
$ \mu_t=m_t$, $\mu_h=\nu_b=m_{\ell\bar{\ell}}$, and $ \mu_s=\nu_s=\mu_{b}=\mu_f= \sqrt{b^2_0/\bT^2+b_0^2/b^2_{\mathrm{max}}}$, with $b_{\mathrm{max}}=0.5\GeV^{-1}$ chosen to prevent parton distribution functions (PDFs) from evolving into the non-perturbative region. 
Each scale is varied by a factor of two, with
\(\mu_b\) and \(\mu_f\) varied both individually and combined.
Second, fixed-order uncertainties are evaluated using the standard seven-point variation of \((\mu_F,\mu_R)\) about the central scale \(m_T\), with \(\mu_F/\mu_R\in[1/2,2]\). Third, matching and non-perturbative uncertainties are assessed from two variations each of the transition profile, of \(c_0\) and of \(\kappa_{\mathrm{NP}}\), within the ranges given in the End Matter and the uncertainties quoted above, with \(B_{\mathrm{NP}}\) held fixed. The uncertainties are combined in quadrature.

Our calculation relies on electroweak parameters defined in the $(G_\mathrm{F},m_Z,\sin^2\theta_W)$ input scheme of Ref.~\cite{Chiesa:2019nqb}, with unstable massive electroweak bosons treated in the complex-mass scheme \cite{Denner:2005fg}. All input masses and couplings are taken from the PDG~\cite{Workman:2022ynf}, and the \texttt{MSHT20an3lo\_as118} PDF~\cite{McGowan:2022nag} is used throughout.

\begin{figure}[!t]
      \centering
      \includegraphics[width=.95\columnwidth]{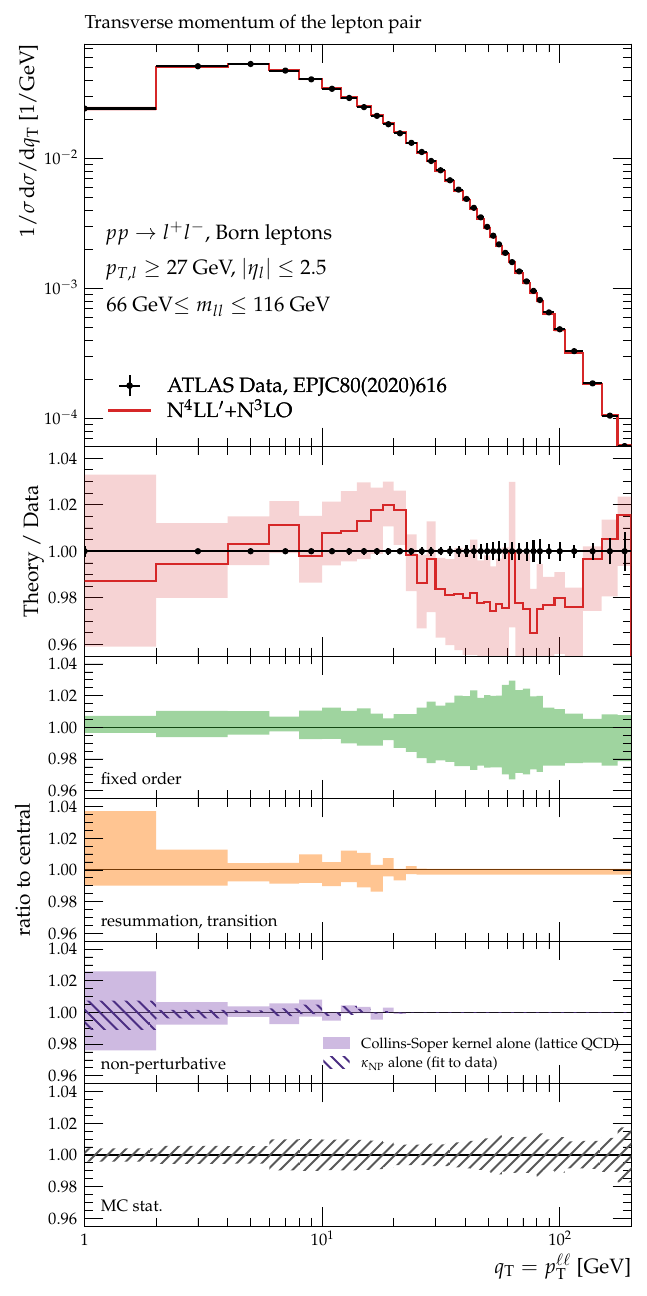}
      \caption{\label{fig:calc}%
      The $\NkLL+\NkLO$ calculation in the ATLAS fiducial phase
      space~\cite{ATLAS:2019zci}, compared to the measured $\qT$ spectrum. 
      Lower panels give the theory uncertainty relative to the central prediction,
      decomposed into fixed-order scales, resummation scales with the transition
      profile, the non-perturbative inputs varied within their quoted uncertainties,
      the Collins--Soper kernel computed in lattice QCD (solid)
      and the width $\kappa_{\mathrm{NP}}$ fitted to data (hatched), and the
      MC integration uncertainty of the fiducial calculation.
      }
\end{figure}
Figure~\ref{fig:calc} shows a comparison between our theory calculation and $13\TeV$ $\qT$ LHC data from ATLAS~\cite{ATLAS:2019zci}. Our 
result agrees with the experimental data within its uncertainty band below
$10\GeV$, where the deviations are at most $1.3\%$; between $10$ and $30\GeV$
they reach $2.0\%$ and above $30\GeV$ $3.5\%$, and the $\chi^2$ per bin against
the theory band and the data uncertainty combined is $0.2$, $1.0$, and $1.6$ in
the three regions. The experimental data are currently two to seventeen times
more precise than the prediction, eight times in a typical bin. In the
first bin, $\qT<2\GeV$, the lattice uncertainty on $c_0$ translates into an
uncertainty of up to $2.6\%$ on the prediction, and of at most $0.8\%$ above $2\GeV$. The fit uncertainty of the width
$\kappa_{\mathrm{NP}}$ moves the prediction by at most $1.1\%$ in the first bin
and $0.5\%$ above it.
The PDF uncertainty has been assessed
in Ref.~\cite{Neumann:2022lft}; it is about a few percent.
The statistical uncertainty of the fiducial calculation
is between $0.4\%$ and $1.7\%$ per bin at ATLAS binning, below $1.5\%$ in all but
the highest bin, and about half the scale uncertainty band in a typical bin and
up to its full size. It is set by computational
resources~\cite{Campbell:2022qmc} and presents a practical bound on the
information content of the calculation.

The calculation of \eq{qTres:mat} is fully differential in the kinematic
variables of the leptons in principle, but the resummation is provided
specifically for two observables: the scaled transverse momentum
$\rT$ and the lepton acoplanarity $\aco$. The information
encoded in the result can be extracted most efficiently in the form
of moments of the distributions~\cite{Assi:2025ibi}, which are
inclusive quantities that are in most cases associated with small
statistical uncertainties. Given the challenges in performing numerically stable
and statistically significant computations at high parametric precision,
the opportunity to use such moments to transfer theory information
to the event simulation presents a significant advantage compared to
binned reweighting techniques that are traditionally used in similar scenarios.

\prlsec{Information-theoretic approach}
An MC generator supplies a prior distribution $q(\Phi)$ over the full
final-state configuration $\Phi$, which is defined in terms of momenta and
particle flavors. To incorporate our theory calculation into the MC generator, we seek the event-level distribution $p(\Phi)$
closest to the prior that reproduces the target moments:
\begin{equation}
  t_j=\frac{1}{\sigma}\int\!\done\sigma\; g_j(\rT,\aco)\,.
  \label{eq:moments}
\end{equation}
Here, each $g_j$ is a product of at most two factors, each a monomial in
$\rT$, $\ln\rT$, $\aco$, or $\ln\aco$ up to the fifth power, for $170$
moments total. Of these, between $14$ and $19$ are eventually chosen by
the algorithm described in the End Matter.
Closeness is measured in terms of the Kullback--Leibler divergence,
\begin{equation}
  L_{\rm KL}(p\|q) = \int\!\df\Phi\, p(\Phi)\,\ln\frac{p(\Phi)}{q(\Phi)}\,
  \label{eq:kl}
\end{equation}
minimized subject to $\int\!\df\Phi\,p\,g_j=t_j$.

The solution to this information-theoretic optimization is a strictly
positive, event-local weight~\cite{Assi:2025ibi,Assi:2026tfh}:
\begin{equation}
  \begin{split}
  p(\Phi)=&\;q(\Phi)\,w(\Phi)\,,\\
  w(\Phi)=&\;\frac{1}{Z(\vec\lambda)}\exp\!\Bigg[\sum_j \lambda_j\, g_j\big(\rT,\aco\big)\Bigg]\,,
  \label{eq:weight}
  \end{split}
\end{equation}
where $\vec \lambda=(\lambda_j)$ are Lagrange multipliers and $Z(\vec\lambda)=\int\!\df\Phi\,q\,\exp[\sum_j\lambda_j g_j]$ enforces
$\int p=\int q$. Since Eq.~\eqref{eq:weight} preserves the normalization
of the prior, the total rate of the calculation is treated as additional
information.

Moments should be matched only as well as they are known in terms of parametric
and statistical precision. For Gaussian moment uncertainties with covariance
$\Sigma$, the constrained problem is the penalized dual,
\begin{equation}
  L(\vec\lambda)=\ln Z(\vec\lambda)-\vec\lambda\!\cdot\!\vec t
      +\tfrac12\,\vec\lambda^{\,T}\Sigma\,\vec\lambda\,,
  \label{eq:dual}
\end{equation}
whose minimum reproduces each moment to within its uncertainty, while
an infinitely uncertain moment decouples. Here, $\vec t=(t_j)$ collects the
target moments. $\Sigma$ is purely statistical and combines the MC
uncertainty of each moment of the calculation with the MC uncertainty
of the same moment in the prior sample. The moments of a set are computed in
one integration, so these uncertainties are correlated. However, we treat them
as independent, to make $\Sigma$ diagonal. This affects only how tightly
each moment is held, and its effect is contained in the out-of-sample agreement
quoted below. The dual is minimized by damped Newton iteration.

For the prior $q(\Phi)$, we use a Sherpa~3~\cite{Sherpa:2024mfk} NLO multi-jet merged
event sample with up to three jets, the first two at NLO and the third at LO
accuracy, with a merging cut of 30~GeV. The jet multiplicity is chosen to
match the perturbative precision of the theory result in the previous section
and is increased to six jets at leading-order accuracy~\cite{Bothmann:2023gew,Bothmann:2023ozs}
for the combined prediction presented in the next section.
The weight of \eq{weight} is applied where our theory calculation is the more accurate
description of the event, which is the case in the region that can be identified
as a zero- or one-jet final state according to the next-to-leading order
multi-jet merging~\cite{Hoeche:2012yf}. Events of higher multiplicity retain
the perturbative input of the multi-jet merged prediction and the associated
theoretical uncertainty. For the calculation at hand they provide equal or
higher perturbative accuracy than the inclusive N$^3$LO theory input.
Approximate electroweak corrections are supplied by \textsc{OpenLoops}~\cite{Buccioni:2019sur} 
and calculated in the scheme of~\cite{Kallweit:2015dum,Lindert:2017olm,Bothmann:2021led}. 
They are applied multiplicatively to the reweighted event sample.
Multiple-parton interactions and beam remnants are simulated
according to~\cite{Pabst:2026pwv}, analyses performed with Rivet~\cite{Bierlich:2019rhm}.

\prlsec{Numerical predictions}
\begin{figure*}[t]
  \centerline{
  \includegraphics[width=.425\textwidth,clip,trim=30 10 0 10]{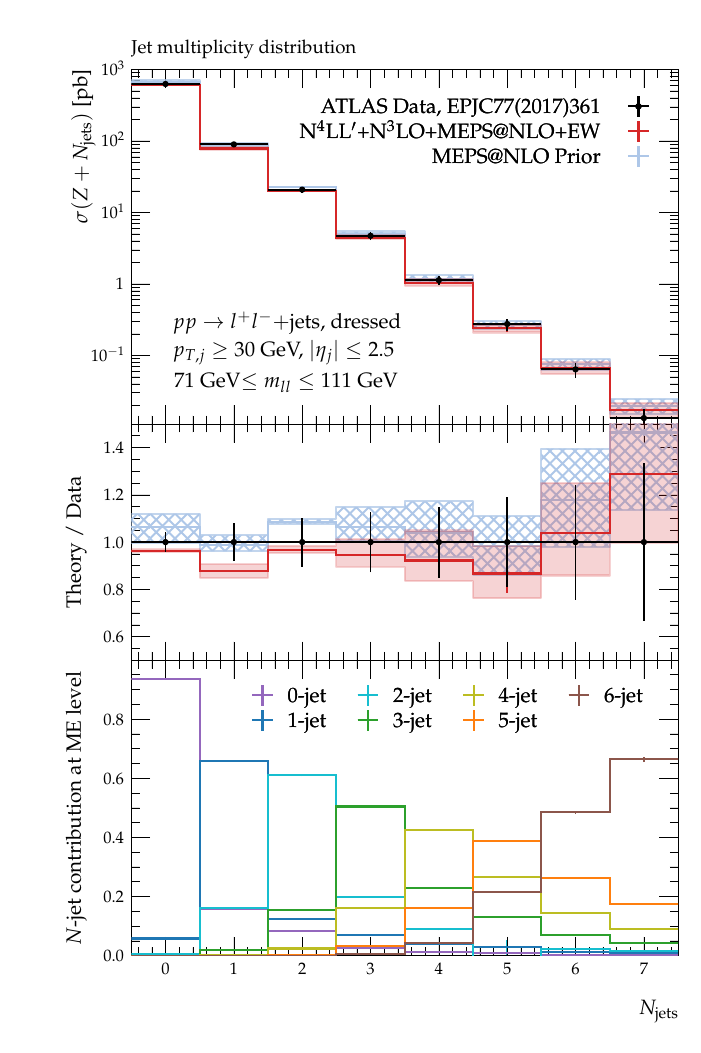}\hspace*{10mm}
  \includegraphics[width=.425\textwidth,clip,trim=25 10 0 10]{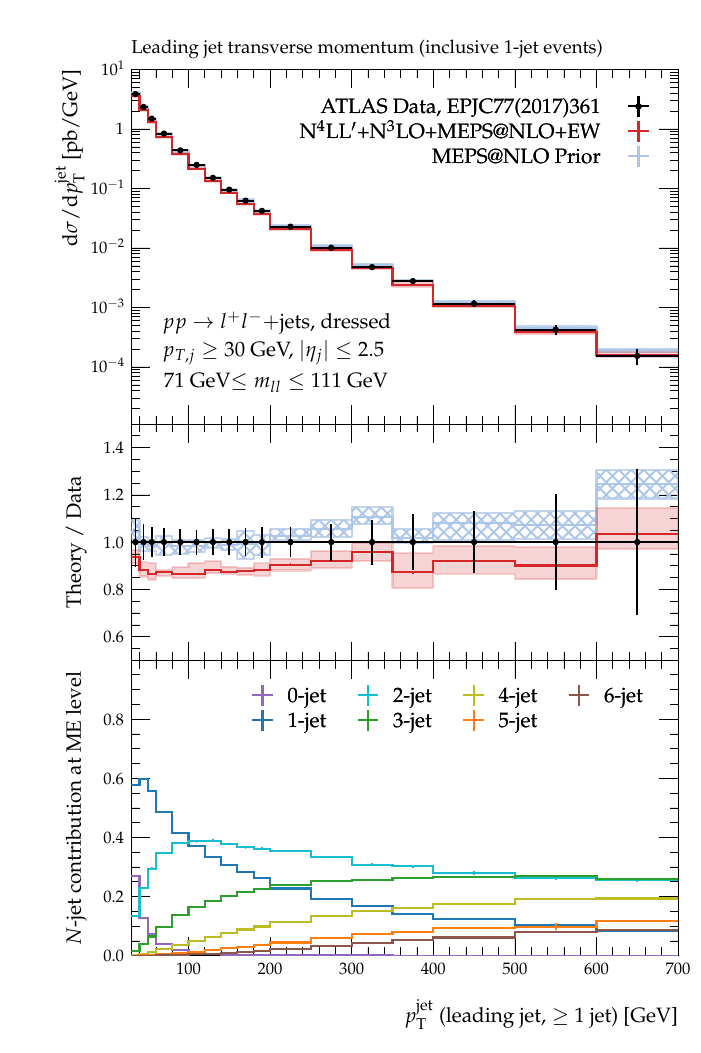}}
  \caption{The NLO multi-jet merged sample at $13\TeV$, before and after
  reweighting with the theory input, compared to LHC experimental data from
  ATLAS~\cite{ATLAS:2017sag}. Left: inclusive jet multiplicity distribution.
  Right: Inclusive transverse momentum distribution of the leading jet.
  The first ratio panel shows the deviation between the simulations and the 
  experimental data. The second ratio panel provides a decomposition of the
  simulated result into sub-samples that have been generated starting from
  different jet multiplicities in the hard process.
  For details on this notation see~\cite{Hoeche:2009rj,Hoeche:2012yf}.}
  \label{fig:events}
\end{figure*}
At $13\TeV$, $15$ moments are enough to reproduce the full distributions.
Applied to the three-jet sample on which the multipliers were derived, the
weights bring the $\rT$ and $\qT$ spectra to within $2\%$ of the calculation
in a typical bin and the acoplanarity to within $0.8\%$, and no bin below the
transition at $\qT=120\GeV$ deviates by more than $7\%$. The prior deviates by
$6\%$ in a typical bin and by up to $40\%$. $20$ of the $32$ bins
in $\rT$ and in $\qT$, and $19$ of $20$ in $\aco$, lie inside the scale band
of the calculation. The
effective sample size is unchanged, $7.6\%$ of the event count before
reweighting and $7.3\%$ after, for this prior with signed weights. The theory
uncertainty of the calculation is propagated through the reweighting (see End
Matter). The same procedure at $\sqrt{s}=13.6\,\TeV$, where no $\qT$ measurement
exists yet, reproduces the calculation to $2\%$ in a typical bin and within
$7\%$ with $14$ moments.

Figure~\ref{fig:events} shows a comparison between a regular MC prediction, our result, and experimental LHC data from the ATLAS collaboration~\cite{ATLAS:2017sag}, for observables that go beyond the $\qT$ and $\aco$ spectra.  This tests the details of the fully exclusive multi-jet event simulation and the impact of the reweighting. Since the reweighting acts only
on $\qT$ and $\aco$, it does not modify jet kinematics, and the description of the exclusive
jet multiplicities in the original multi-jet merged sample is preserved. This is crucial
to correctly model inter-jet correlations at high multiplicity, which are relevant to
most searches for new physics at the LHC. The left panel of Fig.~\ref{fig:events} shows
that beyond 3-jet final states, the uncertainty of the full prediction increases,
in agreement with the fact that only up to three jets are predicted by the precision
theory input, with the third jet at leading-order accuracy.
An additional aspect  is the correct modeling of intra-jet structure at high transverse
momentum. The right panel of Fig.~\ref{fig:events} displays the leading jet $p_T$
spectrum. The second ratio panels shows that perturbative calculations with
two through six final-state jets are most relevant for this observable at high-$p_T$,
all contributing roughly equally to the final result, and indicating that a leading jet
at 700~GeV is likely composed of two to six 30~GeV sub-jets which should be modeled by
dedicated perturbative QCD calculations. An event sample purely based on precision
theory for inclusive observables would not capture the correlations among
these jets correctly, as they would be modeled by a parton shower.

\prlsec{Outlook}
High-precision theory for the lepton-pair transverse momentum and acoplanarity
is now available in a form that experiments can use directly in MC event
simulations. With the Collins--Soper kernel computed from first principles, the
comparison to experimental data becomes a test of QCD in the perturbative and
non-perturbative regime. Our results, and the simulations they enable, provide a sizable
reduction of the theory uncertainty in measurements that rely on the modeling of the
recoil of the Drell--Yan lepton pair. At the same time, fully differential information
in multi-jet final states is retained. Using~\cite{code}, these results can be applied to any MC event generator used as a prior.  Going forward, the same technique could be applied to other high-precision calculations, such as $W$, diboson, and Higgs-boson production.

\begin{acknowledgments}
This manuscript has been authored by Fermi Forward Discovery Group, LLC
under Contract No. 89243024CSC000002 with the U.S. Department of Energy, Office of Science, Office of High Energy Physics.
This research used resources of the National Energy Research Scientific Computing Center (NERSC),
a Department of Energy Office of Science User Facility using NERSC award ERCAP0028985.
This work used the DiRAC Extreme Scaling service (Tursa) at the University of Edinburgh, managed by the EPCC on behalf of the STFC DiRAC HPC Facility. The DiRAC service at Edinburgh was funded by BEIS, UKRI and STFC capital funding and STFC operations grants. DiRAC is part of the UKRI Digital Research Infrastructure.
The work of J.M.C., S.H.\ and M.K.\ was supported by the U.S. Department of Energy,
Office of Science, Office of Advanced Scientific Computing Research,
Scientific Discovery through Advanced Computing (SciDAC-5) program,
grant “NeuCol”.
J.T.\ was supported by the National Science Foundation under Cooperative Agreement
PHY-2019786 (The NSF AI Institute for Artificial Intelligence and Fundamental Interactions,
\url{http://iaifi.org/}), by the U.S. Department of Energy Office of High Energy Physics under grant
number DE-SC0012567, and by the Simons Foundation through Investigator grant 929241.
Research at the Perimeter Institute is supported by the Government of
Canada through the Department of Innovation, Science and Industry Canada, and by the
Province of Ontario through the Ministry of Colleges and Universities. 
B.A.\ was supported in part by the Canadian AI Safety Institute Research Program at CIFAR. 
M.S.\ is funded by the Royal Society through a University Research Fellowship
(URF\textbackslash{}R\textbackslash{}231031).
This work was performed in part at the Aspen Center for Physics, with support for B.A. by a grant from the Simons Foundation (1161654,Troyer).
\end{acknowledgments}

\clearpage
\onecolumngrid
\begin{center}\textbf{\large End Matter}\end{center}
\vspace{0.5em}
\twocolumngrid

\prlsec{Details of the theory calculation}
The transition from the resummation to the fixed-order region is controlled
by the profile:
\begin{equation}
f_{\mathrm{tran}}(\rT)=
\begin{cases}
1, & \rT\le r_0-\delta_r,\\[0.6ex]
1-\dfrac{(\rT-r_0+\delta_r)^2}{2\delta_r^2},
& r_0-\delta_r<\rT\le r_0,\\[1.2ex]
\dfrac{(\rT-r_0-\delta_r)^2}{2\delta_r^2},
& r_0<\rT\le r_0+\delta_r,\\[1.2ex]
0, & \rT>r_0+\delta_r.
\end{cases}
\nonumber
\end{equation}
Our central transition parameters are \((r_0,\delta_r)=(0.20,0.10)\), with matching uncertainties estimated by varying \(r_0\in[0.18,0.22]\) and \(\delta_r\in[0.08,0.12]\). 
The fixed-order contribution \(\done\sigma_{\mathrm{f.o.}}\) is assembled in a hybrid setup using \(\mu_F=\mu_R=m_T\), where \(m_T=\sqrt{m_{\ell\bar{\ell}}^2+\qT^2}\). 
Terms through \(\mathcal O(\alpha_s^2)\) are computed for \(\qT\ge1\GeV\) with Sherpa~\cite{Sherpa:2024mfk,Sherpa:2019gpd}
and \OpenLoops~\cite{Buccioni:2019sur,Cascioli:2011va}. 
The \(\mathcal O(\alpha_s^3)\) correction is supplied by a NNLO \(Z+\)jet calculation
implemented in \MCFM~\cite{Boughezal:2015ded,Campbell:2019dru,Neumann:2022lft},
and evaluated for \(\rT\ge0.06\), corresponding to \(\qT\simeq5.5\GeV\) at the \(Z\) pole. For consistency with the resummed calculation, axial sing\-let contributions are included
in the limit $m_t \to \infty$. This construction provides a complete \(\mathrm{N}^3\mathrm{LL}' + \mathrm{N}^3\mathrm{LO}\) prediction and, upon including the estimated four-loop ingredients, an approximate \(\mathrm{N}^4\mathrm{LL}' + \mathrm{N}^3\mathrm{LO}\) result.

\prlsec{Automatic moment selection}
In total, we tabulate $170$ moments from our theory calculation, though fewer than $20$ are used for the reweighting.
For any prior, moments are selected in the four steps of Table~\ref{tab:selection}~\cite{code}. The screening step keeps only moments determined accurately enough to transfer precision. The admission step adds
moments one at a time, in order of increasing degree, on a fixed fit sample
enriched with the tails of both observables; a moment is kept if the sampling determines the
moment at least as precisely as the calculation does, such that no multiplier is set by noise. The stability step
removes moments until the effective sample size
$\Neff=(\sum_i w_i)^2/\sum_i w_i^2$ stays above its floor and the fit
converges. The last step rejects a
weight that is unbounded as $\rT\to0$ or whose normalization shifts under a
variation. A set that describes any distribution worse than the prior is
refused. The thresholds were fixed once. Two other random fit samples change the set
and move the reweighted distributions by less than $2\%$ in every bin at
$13\TeV$ and by up to $4\%$ at $13.6\TeV$.
\begin{table}[t]
  \caption{\label{tab:selection}%
  Steps of the selection, their criteria, and the surviving moments, for the
  Sherpa priors and the POWHEG prior of \fig{altprior}. $\sigma_{\rm p}$:
  spread of the prior along the new direction; $\Delta C$: shift of the
  normalization constant under a variation.}
  \setlength{\tabcolsep}{2pt}\footnotesize
  \begin{ruledtabular}
  \begin{tabular}{llccc}
    & & \multicolumn{2}{c}{Sherpa} & POWHEG \\
    & & $13$ & $13.6$ & $13\TeV$ \\
    \hline
    tabulated & --                                    & 170 & 170 & 170 \\
    screened  & known to $2\%$                        &  61 &  39 &  61 \\
    admitted  & new direction $<0.1\,\sigma_{\rm p}$  &  20 &  19 &  20 \\
    stable    & $\Neff>10^6$                          &  15 &  14 &  20 \\
    bounded   & finite at $\rT\to0$, $|\Delta C|<\tfrac12$ &  15 &  14 &  19 \\
  \end{tabular}
  \end{ruledtabular}
\end{table}

\begin{figure*}[!t]
  \centering
  \includegraphics[width=\textwidth]{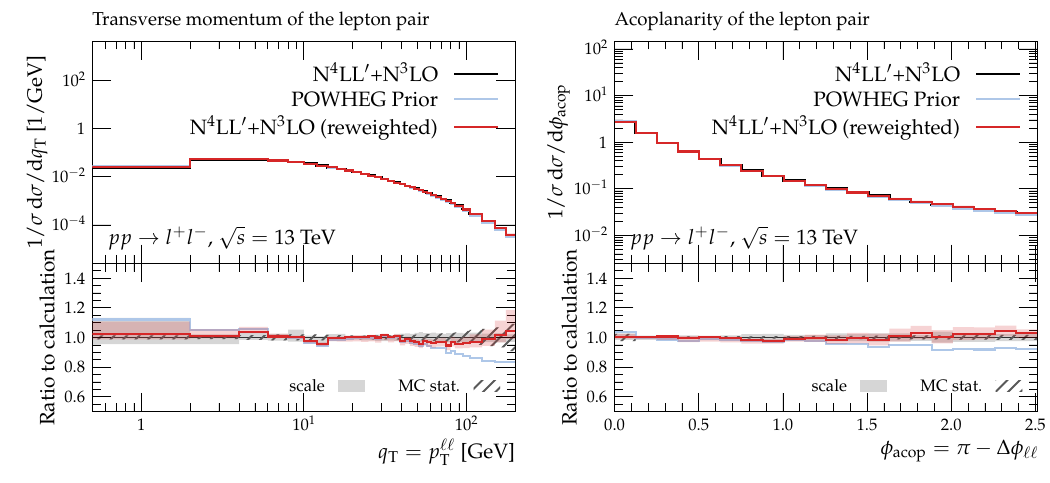}
  \caption{\label{fig:altprior}%
  A POWHEG$+$Pythia~8 sample in the ATLAS $Z\to ee$ production configuration
  (DSID~361106), extended to $\mll>40\GeV$: $\qT$ (left) and acoplanarity
  (right), and their ratio to the $\NkLL+\NkLO$ calculation with its scale
  (grey) and MC (hatched) uncertainties. The prior (light blue)
  departs from the calculation by up to \PWGPRIORDEV. The reweighted sample
  (red, with its propagated theory band) reproduces the calculation to $3\%$
  in a typical bin and within $6\%$ below $\qT=120\GeV$, against $4\%$ and
  $14\%$ for the prior, with the effective sample size unchanged at
  \PWGNEFF{} of $48.6\times10^{6}$ events.
  }
\end{figure*}

\begin{figure*}[!t]
  \centering
  \includegraphics[width=\textwidth]{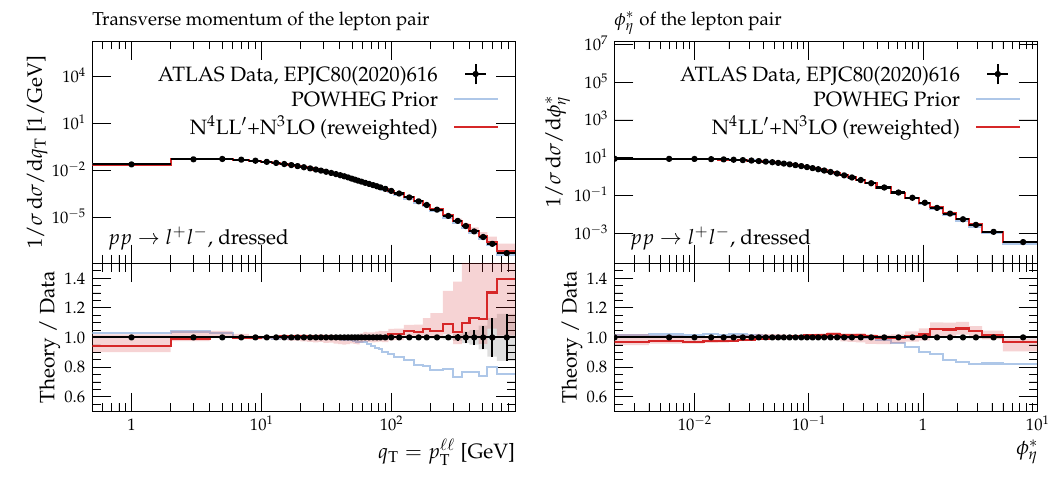}
  \caption{\label{fig:altpriordata}%
  The reweighted POWHEG sample of \fig{altprior} in the fiducial phase space
  of the ATLAS measurement~\cite{ATLAS:2019zci}, $\qT$ (left) and
  $\phistar=\tan(\aco/2)\sin\theta^*_\eta$~\cite{Banfi:2010cf} (right). Below $200\GeV$, the reach of the calculation, the reweighting
  reduces the largest deviation of the $\qT$ spectrum from the data from
  $19\%$ to $6\%$, and the data lie within the propagated theory band
  ($\chi^2$ per bin $0.2$ with the band and the data uncertainty). Between $200$
  and $900\GeV$ the ungated weight extrapolates: the sample
  overshoots the data by $4$ to $40\%$, where the tuned prior undershoots by
  $20$ to $26\%$, within a band that grows to the size of the spectrum. The
  $\phistar$ tail above $1$, where the tuned prior falls $10$ to $18\%$ below
  the data, is corrected to within $6\%$, and below $1$ the reweighted sample
  stays within $3\%$ of the data.
  }
\end{figure*}

\prlsec{Uncertainties and validation}
Uncertainties on the moments transfer to uncertainties on the Lagrange multipliers by linear
response~\cite{Assi:2026tfh}: a shift of the targets, $t\to t+\Delta t$,
shifts the optimum by $\Delta\lambda=H^{-1}\Delta t$ with
\begin{equation}
  H \, =\, \mathrm{Cov}_w(g) \, +\, \Sigma\,.
  \label{eq:hessian}
\end{equation}
The dual Hessian $H$ is evaluated once at the central solution and inverted on the
directions the fit constrains. Each of the $28$ theory variations is propagated as
one correlated shift, and the deviations are combined in quadrature, as for
the calculation itself. The band carries the uncertainty of the calculation
through the admitted moments only: it is narrower than the calculation's in
the first bins of $\qT$ and wider between the peak and the transition to MC,
where the resummation and non-perturbative variations vanish in the calculation
but still move the weight. The absolute scale is set by
$K=\sigma_{\rm calc}/\sigma_{\rm prior}$, delivered for the central prediction
and for each variation.

The distributions of $\rT$, $\aco$, and $\qT$ never enter the fit, only their
moments do, so the agreement quoted in the text is a closure test. The
same pipeline, with the same settings, run on a different prior
POWHEG~\cite{Nason:2004rx,Frixione:2007vw,Alioli:2010xd,Alioli:2008gx}$+$Pythia~8~\cite{Sjostrand:2007gs}
sample in the ATLAS production configuration reproduces the calculation to
$3\%$ in a typical bin~(\fig{altprior}), and
describes the measured $\qT$ spectrum better than the tuned prior up to the
reach of the calculation~(\fig{altpriordata}).

\prlsec{Transition to the prior}
Below the transition to the MC the weight is that of \eq{weight},
above it is the generator's. Specifically, each event carries
\begin{equation}
  w \, =\, K\,\beta(\qT)\, w_0\, e^{\sum_j\lambda_j g_j - C}
     \, +\, \big(1-\beta(\qT)\big)\, w_0\,
  \label{eq:gate}
\end{equation}
with $\beta=1-s(t)$, $s(t)=6t^5-15t^4+10t^3$ and
$t=\mathrm{clip}\big((\qT-120\GeV)/80\GeV,0,1\big)$, the window placed where
the statistical uncertainty of the calculation reaches the size of the
correction it provides. Above it the uncertainty is the seven-point scale
variation of the event sample, blended in the same way; the resummation and
non-perturbative variations vanish there, so the two bands do not double
count. The POWHEG sample does not have a transition to the MC due to its
lack of precision at high jet multiplicity.

\prlsec{Code availability}
The code, the moment selection, the multipliers for the central prediction and
all $28$ variations, the routine that applies \eqs{weight}{gate}, a script
that verifies them, and the figure inputs are available at
\url{https://github.com/benleo12/MomEntER_DY}~\cite{code}.
\clearpage
\bibliography{main}
\end{document}